\documentclass[pdflatex, sn-mathphys-num, iicol]{sn-jnl}% Math and Physical Sciences Numbered Reference Style
\usepackage{graphicx}%
\usepackage{multirow}%
\usepackage{amsmath,amssymb,amsfonts}%
\usepackage{amsthm}%
\usepackage{mathrsfs}%
\usepackage[title]{appendix}%
\usepackage{xcolor}%
\usepackage{textcomp}%
\usepackage{manyfoot}%
\usepackage{booktabs}%
\usepackage{algorithm}%
\usepackage{algorithmicx}%
\usepackage{algpseudocode}%
\usepackage{listings}%
\def\app#1#2{%
  \mathrel{%
    \setbox0=\hbox{$#1\sim$}%
    \setbox2=\hbox{%
      \rlap{\hbox{$#1\propto$}}%
      \lower1.1\ht0\box0%
    }%
    \raise0.25\ht2\box2%
  }%
}
\def\approxprop{\mathpalette\app\relax}

\newcommand{\red}[1]{\color{red}{#1}\color{black}} % remove #1 in brackets to hide comments 
\unnumbered
\title[]{Programmable high-harmonic emission in solids through photon pathways}

\author*[1]{\fnm{Pieter J.} \spfx{van} \sur{Essen}}\email{p.vessen@arcnl.nl}
\author[2]{\fnm{Aday} \sur{Cárdenas}}\email{aday.cardenas@csic.es}
\author[2,3]{\fnm{Rui E.F.} \sur{Silva}}\email{rui.silva@csic.es}
\author[2,3]{\fnm{Álvaro Jiménez} \sur{Galán}}\email{alvaro.jimenez@csic.es}
\author*[1,4]{\fnm{Peter M.} \sur{Kraus}}\email{p.kraus@arcnl.nl}

\affil[1]{\orgname{Advanced Research Center for Nanolithography}, \orgaddress{\street{Science Park 106}, \city{Amsterdam}, \postcode{1098 XG}, \country{The Netherlands}}}
\affil[3]{\orgname{Instituto de Ciencia de Materiales de Madrid, Consejo Superior de Investigaciones Científicas}, \orgaddress{\street{Sor Juana Ines de la Cruz 3}, \city{Madrid}, \postcode{E-28049}, \country{Spain}}}
\affil[2]{\orgname{Max-Born-Institute for Nonlinear Optics and Short Pulse Spectroscopy}, \orgaddress{\street{Max-Born-Strasse 2A}, \city{Berlin}, \postcode{D-12489}, \country{Germany}}}
\affil[4]{\orgdiv{Department of Physics and Astronomy, and LaserLaB}, \orgname{Vrije Universiteit}, \orgaddress{\street{De Boelelaan 1105}, \city{Amsterdam}, \postcode{1081 HV}, \country{The Netherlands}}}

\abstract{
Ultrafast all-optical control of light emission is a central goal of extreme nonlinear optics, with implications for compact short-wavelength sources, petahertz optoelectronics, and label-free super-resolution microscopy. High-harmonic generation in solids is a particularly attractive platform for this goal because it is highly sensitive to both the driving field and the material response, yet a generally applicable framework for controlling harmonic emission has remained elusive.

Here, we demonstrate programmable control of high-harmonic emission in solids and show that it can be quantitatively described within a photon-pathway framework. We find that harmonic emission can be suppressed or enhanced by tuning two experimentally accessible quantities: the effective nonlinear order and the intrinsic emission phase. Across a wide range of semiconductors and dielectrics, this approach unifies parametric and non-parametric modulation, explains distinct delay-dependent spectral responses, and reproduces strong suppression, enhancement, and higher-order pathway revivals. Semiconductor Bloch equation simulations support the model and provide a complementary time-domain picture in which the control field reshapes the interference of sub-cycle emission events.

These results establish high-harmonic generation in solids as a programmable emission process and provide a general route towards ultrafast optical switching, compact coherent short-wavelength sources, and label-free attosecond super-resolution microscopy.
}

\keywords{High-Harmonic Generation, Ultra-Fast Dynamics, Emission Control, Condensed Matter, Attosecond Science}

\begin{document}
\maketitle

\begin{figure*}[htb]
    \centering
    \includegraphics[width=\linewidth]{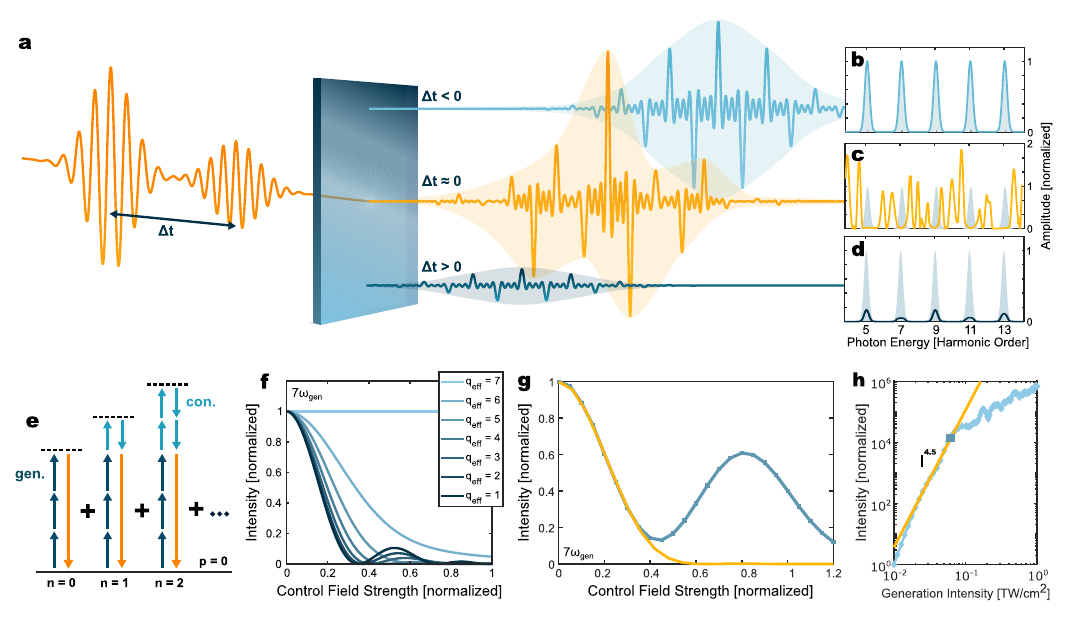}
    \caption{Schematics, calculations, and simulations of harmonic modulation. (a) schematic illustration of harmonic modulation by a two-color field. In (b-c), schematic spectra that can be obtained for the different delays. In (b), the unperturbed harmonic spectra, in (c), the spectra during temporal overlap, and in (d), showing persistent suppression. The shaded areas indicate the unperturbed harmonic spectra. (e) schematically illustrates the photon pathways for the third harmonic. The dark blue arrows correspond to the generation photons, and the light blue arrows to the control photons. All shown transitions are virtual, with the upward arrows corresponding to pumping and the downward arrows to seeding. (f) shows the signal intensities of the seventh harmonic as a function of control intensity for various values of $q_\text{eff}$ calculated using the pathway model.(g,h) show SBE simulation results, which used two bands and a bandgap of 2 eV with a 2000 nm generation pulse modulated by a 1400 nm pulse. (g) shows the simulated normalized intensities of the seventh harmonic for a generation intensity of 50 GW/cm$^2$. The yellow line is the pathway model results for $q_\text{eff} = 4.5$. (h) shows the corresponding intensity dependence of the harmonic on generation intensity. The yellow line indicates an intensity scaling of $q_\text{eff} = 4.5$. }
    \label{fig:pathways}
\end{figure*}

The long-term vision of ultrafast optoelectronics is to process, route, and encode information directly with optical fields on their natural timescale \cite{Schiffrin2013,Ossiander2022,Heide2024,Hassan2024}. This ambition is already driving the emergence of petahertz electronics, in which tailored light waveforms steer charge motion, logic operations, field synthesis, and signal processing far beyond the speed limits of conventional semiconductor technology \cite{Schiffrin2013,Ossiander2022,Heide2024,srivastava2024,zimin2022}. Closely related advances have demonstrated coherent lightwave control of magnetism \cite{Siegrist2019} and vectorial optoelectronic functionality in semiconductors \cite{Sederberg2020}, highlighting that fundamental material observables can now be manipulated on optical-cycle timescales. An equally important, but far less developed, capability is the ultrafast optical control of light emission itself. Such control would be valuable for optical information processing, compact short-wavelength sources, and contrast mechanisms for microscopy and spectroscopy \cite{Cheng2020,RoscamAbbing2024,Bionta2021,Nie2023,Zhang2024a,Koll2025}. Yet existing optical switching concepts remain largely platform-specific, often relying on engineered photonic structures, resonant absorption, or material-specific carrier dynamics \cite{Hassan2024,Cheng2020}.
Recent work has shown that nonlinear metasurfaces enable ultrafast spatial and directional control of harmonic emission by carrier excitation and modification of specific bound-in-continuum resonances, including dynamically programmable nonlinear beam shaping and near-unity all-optical modulation of harmonic generation \cite{Bijloo2024,Bijloo2026}. These advances suggest that nonlinear emission itself can become a programmable optical degree of freedom.

Optically switchable emission processes, in particular fluorescence, underpin modern microscopy and spectroscopy. Optical super-resolution fundamentally relies on switchable responses to engineer high spatial frequencies and increase contrast beyond the diffraction limit \cite{Hell2007}.  A programmable HHG response could provide analogous structured and switchable illumination concepts without requiring fluorescent labels - absent in most semiconductor and condensed-matter systems - enabling label-free super-resolution and computational imaging schemes in otherwise inaccessible material systems.
As nearly all materials emit high harmonics under strong-field driving \cite{Li2020,Ghimire2019,Goulielmakis2022}, high-harmonic generation (HHG) offers a potentially far more universal platform for all-optical emission control. A growing body of work has demonstrated strong modulation of HHG from solids \cite{vanEssen2024,Wang2024,Bionta2021,Nagai2023,Wang2017,Nie2023,Wang2023,Xu2022,Wang2022,Nishidome2020,Cheng2020,vanderGeest2023}. However, the physical origin of these modulations remains fragmented across materials and excitation conditions, and a generally applicable framework for controlling harmonic emission is still missing.

Here we address this challenge by introducing a quantum-optical photon-pathway framework for programmable control of high-harmonic emission in solids. The central advance is not only that harmonic emission can be modulated, but also that suppression and enhancement can be understood and designed through two experimentally accessible quantities: the effective nonlinear order and the intrinsic emission phase. This framework unifies parametric and non-parametric modulation mechanisms, explains why different materials exhibit qualitatively different delay-dependent responses, and provides a route towards programmable ultrafast optical switching of HHG for spectroscopy, compact all-solid coherent sources, and label-free attosecond super-resolution microscopy. In particular, experiments have already shown that this optical control over the HHG process can be exploited to facilitate super-resolution imaging in niobium dioxide \cite{Murzyn2024,Murzyn2026} and zinc oxide \cite{Nie2024}. Here, this framework is generalized and we demonstrate programmable emission control in a wide range of dielectrics and semiconductors, underpinning that the concept holds, in principle, for any solid.

\section{Photon pathways enable programmable high-harmonic emission in solids}

Optical modulation of HHG is realized by combining a strong generation field with a weaker control pulse and recording the harmonic response as a function of delay (Fig. \ref{fig:pathways}a-d). Two limiting mechanisms contribute in multicolor fields beyond single-color emission. In the non-parametric case, the control pulse transiently modifies the material—for example through carrier excitation, phonons, or phase transitions—and can therefore affect HHG beyond temporal overlap (Fig. \ref{fig:pathways}a,d) \cite{deKeijzer2024,Heide2022a,vanderGeest2023,Zhang2024a,Bionta2021,Nie2023,Heide2022b,Wang2024}. A dominant contribution is hot-carrier-induced electron–hole dephasing \cite{Heide2022a,vanderGeest2023,deKeijzer2024}, which damps the coherent polarization and suppresses harmonic emission. In the parametric case, the control field acts directly through the instantaneous electric field during HHG and does not require energy transfer to the material (Fig. \ref{fig:pathways}a,c).

To describe this parametric regime, we introduce a photon-pathway framework for multicolor HHG in solids, adapted from non-collinear gas-phase wavemixing \cite{Vimal2023}. In this picture (see Methods, sec. "Photon Pathway Model"), each harmonic or wavemixing signal arises from a coherent sum of pathways $n$ that differ by the absorption and emission of $n$ control photons each during the light–matter interaction (Fig. \ref{fig:pathways}e). For incommensurate fields, different wavemixing orders $p$ (e.g. $2,+1$ for $p=1$ and $4,-1$ for $p=-1$, see Ext. Data Fig. \ref{fig:ExtData_wavemixing}) are spectrally distinct, while degenerate pathways $n$ contribute to the same emitted frequency. Their relative weights are governed primarily by the effective nonlinear order $q_\text{eff}$.

A central result is that consecutive degenerate pathways contribute with opposite phase. Each photon exchange corresponds to a dipole transition with phase $-i=e^{-i\pi/2}$, such that adding two control photons introduces a phase shift of $-\pi$ \cite{Vimal2023,Gerry_Knight_2004}. A weak control field therefore induces destructive interference with the unperturbed pathway and suppresses the harmonic yield. The strength of this suppression is set by $q_\text{eff}$: perturbative responses remain largely unchanged ($q_\text{eff}=q=7$ in Fig. \ref{fig:pathways}f), whereas non-perturbative responses ($q_\text{eff}\neq q$) show pronounced suppression. At higher control strength, higher-order pathways ($n\geq2$) re-enter with opposite parity and restore constructive interference, producing revivals and enhancement (Fig. \ref{fig:pathways}f). The same interference picture describes the wavemixing orders (Ext. Data Fig. \ref{fig:ExtData_wavemixing}). This establishes HHG modulation as a predictive control problem governed by emission phase and nonlinear order.
More generally, the pathway framework turns solid-state HHG into a programmable interference process, in which optical fields dynamically select, suppress, or revive emission channels on sub-cycle timescales.

Semiconductor Wannier equation (SWE) simulations \cite{Molinero2025} support this picture. For a two-band model driven by a 2000 nm generation pulse and a 1400 nm control pulse, the seventh harmonic exhibits the same progression from suppression to revival (Fig. \ref{fig:pathways}g,h). At low control strength, the pathway model agrees quantitatively when using the simulated $q_\text{eff}$. At higher control strengths, stronger revivals appear, indicating control-induced modification of the nonlinear response beyond the minimal pathway model.

The pathway picture also admits a time-domain interpretation. For incommensurate frequencies, neighbouring half cycles are modulated differently, so the control field alters their relative emission phase. Suppression occurs when half-cycle contributions interfere destructively; revivals occur when constructive interference is restored. In this sense, the modulation acts as temporal phase matching, analogous to Maker’s fringes \cite{Maker1962,Herman1995}, but across optical cycles rather than propagation distance. This interpretation is supported by Gabor analysis of the simulated current, which shows enhanced current amplitude together with disrupted phase relations between neighbouring half cycles (Ext. Data Fig. \ref{fig:ExtData_Gabor}).

Taken together, Fig. \ref{fig:pathways} establishes high-harmonic modulation as a controllable interference problem. A secondary field can selectively suppress or revive harmonic emission by redistributing the interference between degenerate pathways, with the degree of control governed primarily by the effective nonlinear order and the emission phase. The SWE simulations confirm that this mechanism produces strong suppression and revival in a minimal solid-state model, anticipating the near-complete experimental suppression and material-dependent enhancement that is demonstrated below. This provides a predictive framework for designing ultrafast optical control of HHG in solids.

\begin{figure}[htb]
    \centering
    \includegraphics[width=\linewidth]{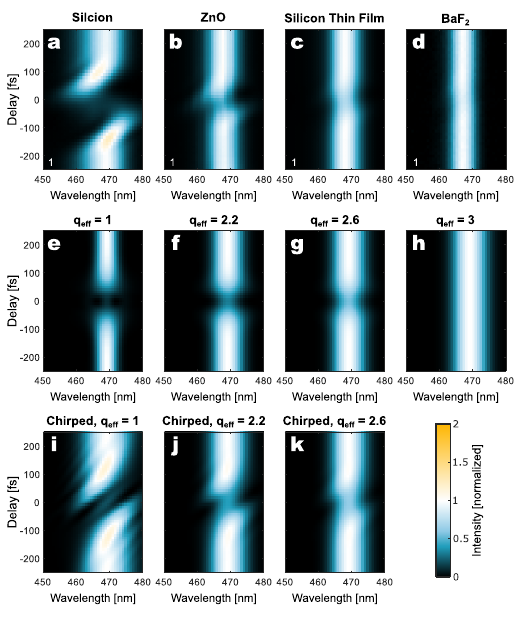}
    \caption{
    (a-d) measured spectra of the third harmonic generated by the 1400 nm beam and modulated by the 2000 nm control beam for various samples. The relative control field strength here is about $1.0\pm 0.1$ with an approximate generation intensity of 40 GW/cm$^2$. (e-k) calculations using the photon pathway model for the different values of $q_\text{eff}$. For (e-h) without including the chirp in the generation beam ($\tau_{1400} = 73$ fs), and (i-k) with inclusion of the chirp ($\tau_{1400} = 115$ fs). The same control pulse is used for all calculations with a comparatively small chirp resulting in a pulse duration of $\tau_{2000} = 47$ fs.}
    \label{fig_materialOverview}
\end{figure}

\begin{figure}[htb]
    \centering
    \includegraphics[width=\linewidth]{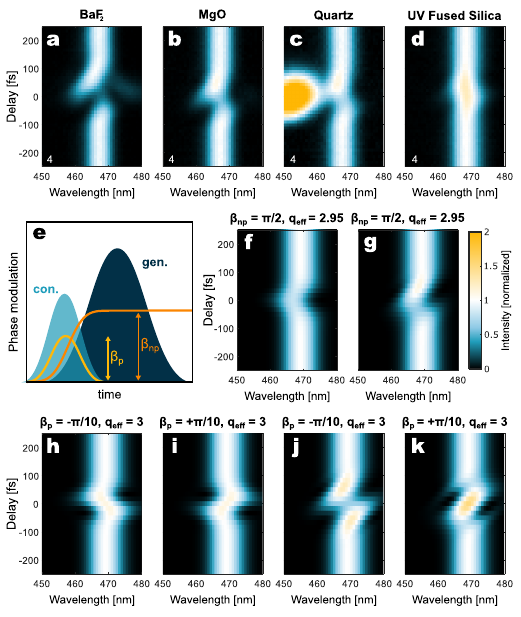}
    \caption{(a-d) measured spectra of the third harmonic generated by the 1400 nm beam and modulated by the 2000 nm control beam for the different high-bandgap semiconductor samples. The relative control field strength here is about $4 \pm 0.4$ with an approximate generation intensity of 40 GW/cm$^2$ and a control field strength of 580 GW/cm$^2$. (e) illustrates the two potential phase modulations induced by the control beam, in yellow the parametric modulation ($\beta_\text{p}$) and in orange the non-parametric modulation ($\beta_\text{np}$). The shaded areas indicate the intensity envelopes of the control and generation pulses. (f,g) calculated spectra illustrating the effect of the non-parametric phase modulation, with (f) without including the chirp of the control beam and (g) with inclusion of the chirp. (h-k) calculated spectra illustrating the effect of the parametric chirp modulation, with (h, i) without including the chirp of the control beam and (j,k) with inclusion of the chirp. The pulse durations are the same as those used for the calculations shown in Fig. \ref{fig_materialOverview}.}
    \label{fig:PhaseOverview}
\end{figure}

\section{A unified description of material-dependent high-harmonic control}

%We perform time-resolved HHG measurements using a fixed 2000 nm control beam and a tunable infrared generation beam in a range of materials (Si, ZnO, MgO, BaF$_2$, quartz, UVFS; see Methods). We first consider the third harmonic generated at 1400 nm (H3-1400).

We next demonstrate this control experimentally in time-resolved HHG measurements for a range of solids. Already at moderate control field strengths, the third harmonic of 1400~nm in silicon is strongly suppressed by a 2000~nm control pulse (Fig. \ref{fig_materialOverview}a), providing an experimental realization of the pathway-interference control introduced above. In contrast, other materials exhibit weaker suppression or qualitatively different responses under identical conditions (Fig. \ref{fig_materialOverview}b-d), showing that the achievable degree and character of control are strongly material dependent.

At low control field strengths (Fig. \ref{fig_materialOverview}a–d), a clear material dependence emerges: low-bandgap materials (Si, ZnO) show pronounced suppression, whereas high-bandgap materials show little to no modulation. Independent intensity-scaling measurements reveal that Si and ZnO are non-perturbative, while the remaining materials follow perturbative scaling (Ext. Data Fig. \ref{fig:ExtData_IntensityScalings}).

Using these measured scalings as input, the photon-pathway model reproduces the observed suppression across materials through a single effective parameter $q_\text{eff}$ (Fig. \ref{fig_materialOverview}e–h). This is a central result: the achievable degree of control is largely captured by the effective nonlinear order, which compactly describes the relevant material- and excitation-dependent nonlinear response.

Beyond yield suppression, the experiments also show pronounced anti-symmetric spectral shifts, most clearly in Si and ZnO. These are reproduced by including the measured pulse chirp (Fig. \ref{fig_materialOverview}i–k). The shifts arise because, during partial pulse overlap, different spectral components are suppressed unequally. This also explains the weak enhancement observed around $\pm 100$ fs in Si.

Taken together, these results show that material-dependent HHG modulation can be predicted within a unified and minimal framework: $q_\text{eff}$ governs the strength of suppression, while the temporal field profile governs the spectral reshaping. This identifies experimentally accessible programming parameters for HHG modulation across a wide range of solids.

\section{The role of the emission phase in modulation of high-harmonic generation}
\begin{figure*}[htb]
    \centering
    \includegraphics[width=\linewidth]{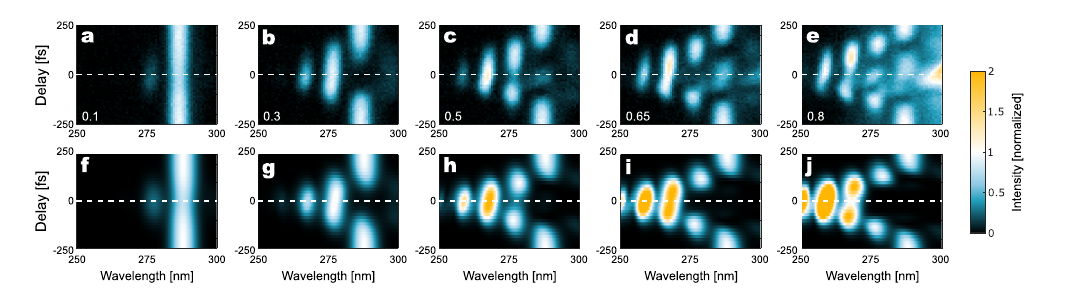}
    \caption{Experimental (a-e) and pathway model (f-j) results of H7 generated from ZnO with a 2000 nm generation beam at 300 GW/cm$^2$ and a 1600 nm control beam. From (a-e) and (f-j) for increasing control field strengths of respectively 0.1, 0.3, 0.5, 0.65, and 0.8. The white dashed line indicated a delay of 0 fs. The pathways model uses $q_\text{eff} = 2.5$. The calculations used pulse durations of $\tau_{2000} = 41$ fs and $\tau_{1600} = 154$ fs.}
    \label{fig:H7_overview}
\end{figure*}
\begin{figure}[htb]
    \centering
    \includegraphics[width=\linewidth]{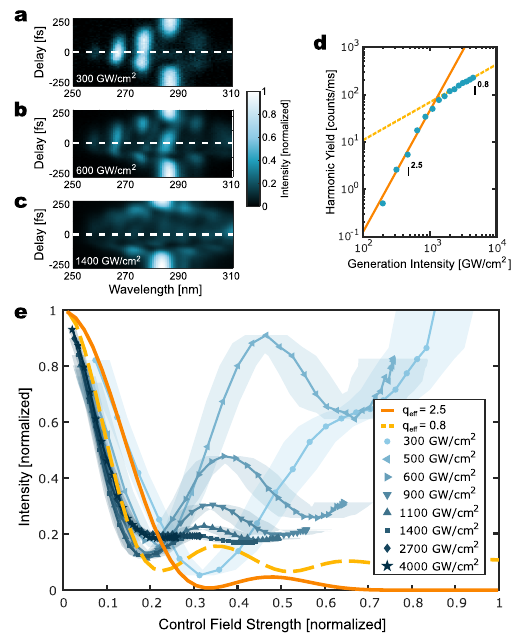}
    \caption{Experimental results of H7 generated from ZnO with a 2000 nm generation and a 1600 nm control beam. (a), (b), and (c)
    show the delay-spectra for a relative control field strength of 0.4 for various generation intensities of respectively 300 GW/cm$^2$, 600 GW/cm$^2$, and 1400 GW/cm$^2$. The white dashed line indicates the delay of 0 fs.
    (d) shows the intensity scaling of H7 where the solid orange line corresponds to $q_\text{eff} =2.5$ and the dashed yellow line to $q_\text{eff} =0.8$.(e) shows the H7 harmonic intensity at the delay of 0 fs as a function of control field strength for the various generation intensities. The shaded area indicates the uncertainty in control field strength; the uncertainty in intensity is not shown here as it is small in comparison. The solid orange and dashed yellow line indicates the pathway model results for respectively $q_\text{eff} =2.5$ and 0.8, taking into account both the spatial and temporal profile of the beams. For the yellow line, which considers the higher total intensity, we also consider the intensity-dependent emission phase contribution with $\beta_\text{p} = \pi$.} 
    \label{fig:H7_closeup}
\end{figure}
\begin{figure*}[htb]
    \centering
    \includegraphics[width=\linewidth]{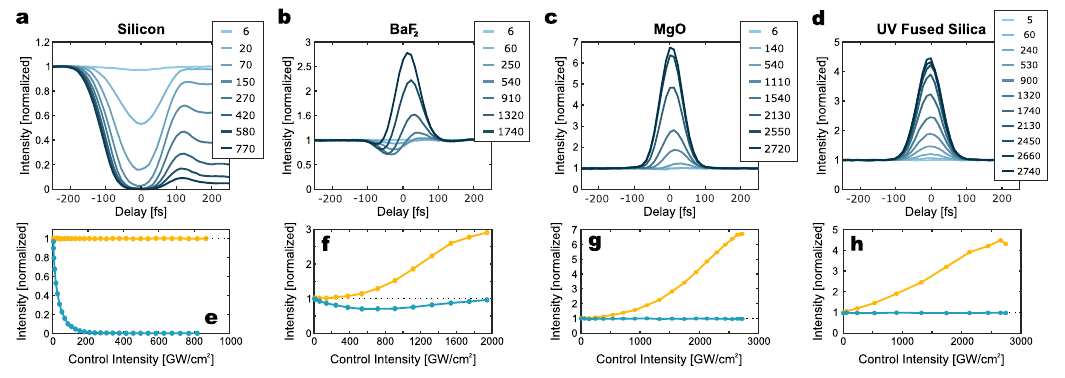}
    \caption{Third harmonic intensity generated by 1400 nm and modulated by the 2000 nm control beam from different materials. (a-c) show the integrated harmonic yield as a function of delay for various control field strengths. The legend indicated the control field strength in GW/cm$^2$ while the generation intensity was fixed at approximately 40 GW/cm$^2$. (e-h) show the maximum enhancement in yellow, and the maximum suppression in blue.}
    \label{fig:IntensityModulation}
\end{figure*}

At higher control intensities, pronounced dynamics also emerge in high-bandgap materials. For BaF$_2$, MgO, and quartz, we observe asymmetric spectral shifts accompanied by asymmetric enhancement and suppression around temporal overlap (Fig. \ref{fig:PhaseOverview}a–c). In quartz, an additional even-order wavemixing contribution appears, consistent with its broken inversion symmetry. In contrast, UVFS exhibits predominantly anti-symmetric enhancement without suppression.

We attribute these distinct behaviors to modulation of the emission phase $\phi_e$ over the generation pulse. Within our framework, two contributions can be distinguished: a parametric phase, arising from the instantaneous field and scaling with control intensity, and a non-parametric phase, arising from transient material excitation and scaling with the accumulated control-pulse intensity. These contributions are schematically illustrated in Fig. \ref{fig:PhaseOverview}e.

A key observation is that these two phase channels produce qualitatively different signatures. Non-parametric phase modulation leads to symmetric spectral shifts around temporal overlap, while parametric phase modulation preserves anti-symmetric features. The experimental data for BaF$_2$, MgO, and quartz require a non-parametric contribution, whereas the anti-symmetric response in UVFS indicates a dominant parametric phase.

Quantitative agreement is obtained by combining both phase contributions in the pathway model. The resulting spectra reproduce the observed asymmetries and regions of enhancement and suppression (Fig. \ref{fig:PhaseOverview}f–k). While the temporal structure of the driving fields influences the detailed spectral shape, the dominant effect is the phase modulation itself: the control field modifies the relative phase between sub-cycle emission events, thereby altering their interference. In this sense, the modulation can be interpreted as temporal phase matching.

Importantly, the symmetry of the modulation provides a direct experimental handle to disentangle the underlying mechanisms. Anti-symmetric responses indicate parametric phase modulation, whereas the breakdown of this symmetry signals non-parametric contributions. This enables a separation of field-driven and material-driven effects without requiring microscopic modeling.

We note that similar spectral features have been reported previously in quartz \cite{Zhang2025}. Here, our photon-pathway framework provides a complementary interpretation: Our model fully captures the results from ref. \cite{Zhang2025}, where the experimental results are rationalized as band-gap modulation during pulse overlap. Our model can additionally describe the observed center-frequency modulation of HHG as the result of any parametric phase evolution (besides a transient band-gap shift, this could, for instance, be cross-phase modulation). Interestingly, our framework helps identify the observed asymmetry of the HHG signal before and after time zero, which occurs both in ref. \cite{Zhang2025} and Fig. \ref{fig:PhaseOverview}a–c, as a signature of competing parametric and non-parametric phase contributions.

\section{Higher order pathways in high-harmonic generation from ZnO}
To illustrate the applicability of the framework beyond the third harmonic, we now consider the seventh harmonic generated at 2000 nm (H7-2000) and modulated by a 1600 nm control field in ZnO. ZnO combines a high damage threshold with clearly non-perturbative intensity scalings, making it an ideal system to access higher-order pathway dynamics. Time-resolved measurements for increasing control strength are shown in Fig. \ref{fig:H7_overview}a-e. At low control powers, H7 is suppressed, analogous to H3-1400, while prominent spectral shifting is absent because the 2000 nm generation pulse is only weakly chirped. 

At the same time, the spectra directly reveal the redistribution of emission into wavemixing orders. In Fig. \ref{fig:H7_overview}a, H7 ($p=0$) is accompanied by sum-frequency wavemixing signals, most prominently $p=1$ ($6,+1$) and $p=2$ ($5,+2$), while a difference-frequency contribution $p=-1$ ($8,-1$) also appears. Their emergence coincides with depletion of H7 and reflects the first degenerate pathway contribution to the harmonic, $p=0,n=1$, which exchanges two additional control photons and interferes destructively with the unperturbed pathway. At higher control strength, a revival of H7 becomes apparent in Fig. \ref{fig:H7_overview}c. Within the pathway picture, this marks the increasing contribution of the next degenerate pathway, $p=0,n=2$, which exchanges four additional control photons and is therefore again in phase with the unperturbed emission. Correspondingly, the wavemixing signals are reduced. At still higher control strengths (Fig.~\ref{fig:H7_overview}d,e), revivals also emerge in the wavemixing channels, consistent with the same interference mechanism acting across the full set of pathways.

The corresponding photon-pathway calculations are shown in Fig. \ref{fig:H7_overview}f-j. At low control intensities, the model reproduces the experimental results remarkably well, for both the harmonic and the wavemixing orders. At higher intensities, it continues to capture the onset and phase of the revivals, but underestimates their strength. This already indicates that higher-order pathways contribute more strongly than predicted by the minimal analytical model.

We performed the same measurements for different generation intensities, with similar behavior shown in Fig. \ref{fig:H7_closeup}a-c. As the generation intensity increases, suppression and revival shift to lower relative control intensities, and difference-frequency generation becomes more prominent. For the highest generation intensity (Fig. \ref{fig:H7_closeup}c), the harmonic additionally blue-shifts, indicating a non-parametric emission-phase modulation induced by the generation pulse itself, most likely through carrier excitation. Consistently, the measured intensity scaling of H7-2000 in ZnO (Fig. \ref{fig:H7_closeup}d) yields two distinct effective nonlinear orders in the low- and high-intensity regimes, 2.5 and 0.8, respectively, explaining the earlier onset of suppression at higher generation intensity.

For a quantitative comparison, we focus on the harmonic intensity at zero delay (Fig. \ref{fig:H7_closeup}e). The pathway model reproduces the observed suppression rates through the measured effective nonlinear orders and also captures the phase of the revival. The remaining discrepancy is systematic: the higher-order pathways contribute more strongly in the experiment than in the analytical model. The deviation depends on both the absolute intensity and the control wavelength (Ext. Data Fig. \ref{fig:ExtData_WavelengthDependence}), showing that these higher-order contributions are themselves strongly tunable. 

This behavior is consistent with the breakdown of the assumption that the control field leaves $q_\text{eff}$ unchanged at high fields, as also noted in Ref. \cite{Vimal2023}. In addition, higher-order pathways are expected to be enhanced by resonant transitions. At the photon energies used here, such transitions lie within the joint density of states between valence and conduction bands, which increases their oscillator strength. Indeed, allowing for a phenomenological renormalization of the higher-order pathway amplitudes restores quantitative agreement with experiment (see Supplementary Material, Fig. 5). This indicates that higher-order pathways contribute more strongly than predicted by the minimal analytical expansion, rather than pointing to a different underlying mechanism.

 \section{High-harmonic modulation as programmable ultra-fast all-optical switch} 

Figure \ref{fig:IntensityModulation} summarizes the achievable control over high-harmonic emission. Panels (a,e) show strong suppression (Si), while panels (b–d,f–h) demonstrate enhancement in other materials. The degree of control is substantial: in silicon, the characteristic suppression intensity ($I_\mathrm{sat}$, defined as 50\% reduction) is $\sim22$~GW/cm$^2$, while the highest applied control intensity reaches $\sim816$~GW/cm$^2$, close to the damage threshold. This corresponds to a ratio of $\sim37$ between the damage threshold and the saturation intensity, sufficient for an estimated $\sim6\times$ resolution improvement in HADES-type microscopy \cite{Murzyn2024}. At higher control intensities ($\gtrsim400$~GW/cm$^2$), suppression becomes complete within the experimental signal-to-noise ratio, corresponding to a reduction of the harmonic yield by more than three orders of magnitude.

The observed behavior follows a simple control principle. Parametric pathway interference provides efficient suppression when the effective nonlinear order differs strongly from the perturbative order ($q_\mathrm{eff}\ll q$), while enhancement occurs when $q_\mathrm{eff}\approx q$, such that higher-order pathways are weak and the influence of the control pulse on the emission phase dominates. In this regime, the control field modifies the emission phase and improves the temporal coherence of the harmonic buildup, effectively compensating the intrinsic intensity-dependent dipole phase mismatch between adjacent half cycles \cite{2025Kuzkova}.

In contrast, strong suppression arises when large differences between $q_\mathrm{eff}$ and $q$ enable destructive interference between degenerate pathways. Purely parametric suppression, however, is intrinsically limited by higher-order pathways, which lead to revivals at elevated control intensities. In the present experiments, these revivals are efficiently suppressed by non-parametric processes, in particular carrier-induced dephasing.

This is illustrated in Fig. \ref{fig:IntensityModulation}a,e for silicon. At low control intensities, suppression is symmetric around temporal overlap and governed by parametric pathway interference. At higher intensities, non-parametric effects dominate, leading to persistent suppression beyond temporal overlap and the loss of symmetry. Crucially, this also suppresses pathway-induced revivals, enabling robust and near-complete switching.

Carrier-induced dephasing provides a natural mechanism for this stabilization. In our previous work, the harmonic yield was shown to scale approximately as 
$I \approxprop e^{-\Delta t/T_2}\left|T_2\left(1-e^{-\Delta t/T_2}\right)\right|^2$, 
where $T_2$ is the coherence time and $\Delta t$ the excursion time. As $T_2$ decreases with increasing carrier density due to enhanced scattering, even modest reductions in $T_2$ lead to a strong, effectively exponential suppression of HHG emission. This provides a powerful non-parametric route to stabilize suppression against revivals.

In addition to carrier dynamics, we observe long-lived modulation in BaF$_2$ due to excitation of optical phonons (Ext. Data Fig. \ref{fig:ExtData_Phonons}). The resulting oscillatory modulation of the emission phase produces periodic spectral shifts, illustrating that collective excitations offer an additional handle for programming HHG emission.

The demonstrated control directly translates into ultrafast switching. Because the modulation is confined to the temporal overlap of the pulses, the switching speed is fundamentally limited by the pulse duration. Even for $\sim100$~fs pulses this corresponds to switching rates in the tens of THz range, with the underlying mechanism extending towards sub-cycle, potentially petahertz switching for shorter, carrier-envelope-phase-stable waveforms.

More broadly, the results establish a unified strategy for programming HHG emission across materials. While Fig. \ref{fig:IntensityModulation} compares different systems, the underlying control parameter is the effective nonlinear order $q_\mathrm{eff}$. Enhancement is obtained when $q_\mathrm{eff}\approx q$, whereas strong suppression is obtained when $q_\mathrm{eff}\ll q$. Importantly, $q_\mathrm{eff}$ can be tuned not only by material choice but also by excitation conditions—for example by selecting wavelengths that drive (higher) harmonics above the bandgap, where non-perturbative scalings typically emerge. This provides a practical route to switch between enhancement and suppression within a single system.

Taken together, these results demonstrate that high-harmonic generation in solids can be programmed with high contrast and ultrafast speed using purely optical means. More broadly, they establish HHG as a programmable emission process, in which the interplay of nonlinear order, emission phase, and pathway interference can be used to design light on optical-cycle timescales. In this context, optically programming emission itself complements recent advances in petahertz electronics, extending ultrafast control from charge dynamics to the emitted field. The resulting framework provides a general route toward ultrafast switching, novel contrast mechanisms, and tailored short-wavelength sources in solid-state platforms.

% \section{Online content}
% Any methods, additional references, Nature Portfolio reporting summaries, source data, extended data, supplementary information, acknowledgements, peer review information; details of author contributions and competing interests; and statements of data and code availability are available at \red{Insert Link by Editor}.

\section*{Methods}\label{sec11}
\bmhead{Experimental Setup}
We used a Light Conversion ORPHEUS-MIR optical parametric amplifier (OPA) pumped by an 80 W CARBIDE laser at a repetition rate of 199.7 kHz. The fixed 2000 nm output together with a tunable output (1350 nm to 2000 nm) were used as generation and control beams. None of the used OPA outputs had carrier-envelope phase (CEP) stability. 
A schematic of the setup can be found in Ext. Data Fig. \ref{fig:ExtData_Schematic}. We use a half-wave plate (HWP) and polarizer combination in both beam paths to attenuate the beams and to fixed the polarization to be parallel and linear (S-polarized with respect to the mirrors). The telescopes in both arms are used to control the divergence and relative spot sizes of the beams, with the telescope lenses being calcium fluoride. A motorized delay stage enabled scanning the delay between the pulses while a manual stage was used for coarse alignment. A 50/50 beamsplitter (designed for 2000 nm) was used to combine the beams collinearly. The beams were weakly focused onto the sample using a 100 mm calcium fluoride lens and collected in transmission using a 50 mm calcium fluoride lens. To separate the harmonic emission from the infrared drivers we made use of a set of UV transmission gratings. In addition to separating the drivers from the harmonics, the gratings allow us to tune the detection such that we can detect the higher harmonics without saturating the detectors with the lower harmonics. The downside of this technique is that we have a non-flat spectral response of our detector, inhibiting the direct comparison of the yield of the different harmonics and wavemixing signals. The UV gratings limit our detection to a minimum of 250 nm. Both the harmonic and infrared signals were focused into multimode fibers, which were connected to the spectrometers. The harmonic spectra are detected using a QEpro Ocean Optics spectrometer, while the IR is detected using an AvaSpec-NIR spectrometer. After the gratings in transmission a silicon plate with a high reflection coating designed for the harmonic wavelengths of 2000 nm allowed us to image the harmonic spots onto a monochrome CMOS camera which we used for alignment.\\ 

The pump-probe scheme employed here has been used in numerous HHG experiments. However, there is a clear distinction to be made between colinear and non-colinear geometries as well as between commensurate and incommensurate wavelength combinations. By introducing an angle between the control and generation beam, as employed by Refs. \cite{Vimal2023, Luttmann2023, Zhang2025, RoscamAbbing2024, Bertrand2011, Heyl2014}, the harmonics and wavemixing orders get spatially separated. In our experiments, we instead spectrally separate the harmonics and wavemixing by having incommensurate frequencies. We note that these two cases respectively exploit the conservation of momentum and energy. In the case where the wavemixing and harmonics are neither spatially nor spectrally separated, the observed dynamics are dominated by the interference of the different photon pathways, as different photon combinations can yield the same energy, rather than the Maker's fringe-type behavior observed in our experiments. This is observed in collinear $\omega+2\omega$ \cite{He2010, Burger2017} and $\omega+3\omega$ \cite{Burger2017,Kroh2018} experiments as well as RABBIT experiments \cite{Paul2001}.\\

We employ weak focusing to have a uniform intensity profile within the samples, in order to minimize changes to the phasematching. The relative spot size and pulse duration of the control and generation beam can greatly impact the observed modulation. In particular, when the control beam is smaller and shorter than the generation beam, the interaction is restricted to a limited time and space, thus limiting the amplitude of the modulation that can be observed. This becomes particularly important when multiple effects contribute to the modulation.  For example, while the pathway model depends on the relative field strength, excitation-induced dephasing requires multi-photon excitation, which, depending on the material, might be restricted to only a small area in the high-intensity center of the control beam. This can thus lead to an inhomogeneous modulation in the radial direction of the control beam, making the observed harmonic modulation dependent on the size of the generation beam. In all our experiments, the control spot size is either similar or greater than the generation spot size. Details on the calibration of the spot size and pulse duration can be found in the Methods section, together with all values for the experimental data shown in this work. \\

Having both the generation and control wavelength in the infrared range minimized the multi-photon absorption rate, enabling us to perform measurements at high field strengths with minimal direct carrier excitation.  \\

\bmhead{Intensity calibration} 
All the experimental intensities mentioned in this work are the pulse-averaged incident vacuum intensities, evaluated using:
\begin{equation}
    I = \frac{P}{A \cdot\tau \cdot\Gamma_{\text{rep}}} \quad \text{ with} \quad A = \pi w_0^2
\end{equation}
Here, $P$ is the total input power, $w_0$ the $e^{-2}$-width of the intensity profile in focus, $\Gamma_\text{rep}$ the repetition time of the laser, and $\tau$ the full-width at half-maximum (FWHM) pulse duration. The reflections of the sample surfaces are ignored.\\

\bmhead{Spot size}
The spot sizes were evaluated using a one-dimensional knife-edge, where the derivative of the measured intensity was fitted using a Gaussian \cite{deAraujo2009}:
\begin{equation}
    \frac{\text{d}I(x)}{\text{d}x} = \frac{P}{w_0\sqrt{\pi}} e^{-\left(\frac{x-x_0}{w_0}\right)^2}
\end{equation}
The measured spot sizes are shown in Table \ref{tab:spotsize}.\\

\begin{table}[htb]
\begin{tabular}{|l|l|l|}
\hline
\textbf{Figure} & \textbf{Wavelength} [nm] & \multicolumn{1}{c|}{$w_0$ [\textmu m]} \\ \hline
\ref{fig_materialOverview}, \ref{fig:PhaseOverview}, \ref{fig:IntensityModulation}, \ref{fig:ExtData_Polarization}, \ref{fig:ExtData_IntensityScalings},\ref{fig:ExtData_Phonons} & 2000             &          $29.4\pm 1.0$                           \\ \hline
\ref{fig_materialOverview}, \ref{fig:PhaseOverview}, \ref{fig:IntensityModulation}, \ref{fig:ExtData_Polarization}, \ref{fig:ExtData_IntensityScalings},\ref{fig:ExtData_Phonons} & 1400             &          $11.9\pm0.3$                        \\ \hline
\ref{fig:H7_overview}, \ref{fig:H7_closeup} & 2000        &          $10.0\pm0.4$                            \\ \hline
\ref{fig:H7_overview}, \ref{fig:H7_closeup} & 1600        &          $11.7\pm0.5$                            \\ \hline
\ref{fig:ExtData_WavelengthDependence} & 2000             &          $10.0\pm 0.4$                             \\ \hline
\ref{fig:ExtData_WavelengthDependence} & 1400             &          $13.9 \pm 0.2$                             \\ \hline
\ref{fig:ExtData_WavelengthDependence} & 1450             &          $12.7 \pm 0.3$                             \\ \hline
\ref{fig:ExtData_WavelengthDependence} & 1500             &          $12.7 \pm 0.4$                             \\ \hline
\ref{fig:ExtData_WavelengthDependence} & 1550             &          $11.3 \pm 0.3$                             \\ \hline
\ref{fig:ExtData_WavelengthDependence} & 1600             &          $12.0 \pm 0.4$                             \\ \hline
\ref{fig:ExtData_WavelengthDependence} & 1650             &          $12.3 \pm 0.4$                             \\ \hline
\ref{fig:ExtData_WavelengthDependence} & 1700             &          $14.3 \pm 0.6$                             \\ \hline
\end{tabular}
\caption{Overview of the measured spot sizes. The indicated errors are the fitting errors.}
\label{tab:spotsize}
\end{table}

\bmhead{Pulse duration}
The pulse duration was extracted from the wavemixing cross-correlation measured in UV fused silica, details on this procedure can be found in the Supplementary Material. The measured pulse durations are shown in Table \ref{tab:pulseduration}. 
\begin{table}[htb]
\begin{tabular}{|l|l|}
\hline
\textbf{Wavelength} [nm] & \multicolumn{1}{c|}{$\tau$ [fs]} \\ \hline
2000             &          $45\pm7$                           \\ \hline
1400             &          $114\pm 4$                           \\ \hline
1450             &          $142\pm4$                            \\ \hline
1500             &          $156\pm5$                            \\ \hline
1550             &          $152\pm5$                             \\ \hline
1600             &          $156\pm6$                             \\ \hline
1650             &          $167\pm5$                             \\ \hline
1700             &          $161\pm4$                             \\ \hline
\end{tabular}
\caption{Overview of the measured pulse durations. The errors include both the fitting errors and the uncertainty due to the finite sample thickness.}
\label{tab:pulseduration}
\end{table}

\bmhead{Sample Characteristics}
The following samples were used in the experiments. 150 \textmu m thick [0 0 1] silicon. 200 nm thick amorphous silicon thin film on a 500 \textmu m thick fused quartz substrate fabricated using chemical vapor deposition. 100 \textmu m thick [0 0 0 1] zinc oxide. 250 \textmu m thick [1 1 0] barium fluoride. 100 \textmu m thick [0 0 1] magnesium oxide. 150 \textmu m thick [0 0 0 1] $\alpha$-quartz. 100 \textmu m amorphous UV fused silica. All samples were two-sided polished. The crystal orientation with respect to the polarization was calibrated by measuring the harmonic polarization dependence in the different samples, as can be found in Ext. Data Fig. \ref{fig:ExtData_Polarization}.

\bmhead{Photon Pathway Model}

To describe parametric two-colour modulation of HHG, we use an analytical photon-pathway model based on the thin-slab approximation. The analytical structure of this pathway expansion follows the framework introduced in Ref. \cite{Vimal2023}, here adapted to collinear incommensurate multicolour HHG in solids. The emitted field of the $q$th harmonic generated by a field $E_\mathrm{g}=|E_\mathrm{g}|e^{i\phi_\mathrm{g}}$ is written as
\begin{equation}
    E_q \propto |E_\mathrm{g}|^{q_\mathrm{eff}} e^{iq\phi_\mathrm{g}+i\phi_e},
    \label{eq:Eq}
\end{equation}
where $q_\mathrm{eff}$ is the effective intensity scaling of the harmonic and $\phi_e$ is the intrinsic emission phase, that is, the phase offset between the emitted harmonic and the driving field. We refer to $\phi_e$ as the emission phase rather than explicitly the dipole phase, because here it is introduced as a temporal phase offset rather than the spectral phase commonly used in HHG \cite{Carlstrm2016,Lewenstein1995}. Equation \ref{eq:Eq} can be interpreted as an effective nonlinear response that generalizes the perturbative scaling $E_q \propto |E_g|^q$ to the non-perturbative regime via $q_\mathrm{eff}$. In this sense, $q_\mathrm{eff}$ captures deviations from simple power-law scaling and provides a compact description of the underlying strong-field dynamics.

In the presence of a weaker control field $E_\mathrm{c}=|E_\mathrm{c}|e^{i\phi_\mathrm{c}}$, the total field becomes $E_\mathrm{tot}=E_\mathrm{g}+E_\mathrm{c}$. Expanding Eq. \ref{eq:Eq} in powers of $|E_\mathrm{c}/E_\mathrm{g}|$ gives
\begin{equation}
\begin{split}
    E_q \propto &\, |E_\mathrm{g}|^{q_\mathrm{eff}} e^{i\phi_e}
    \sum_{p=-\infty}^{\infty} e^{i(q-p)\phi_\mathrm{g}+ip\phi_\mathrm{c}}
    \sum_{n=0}^{\infty}\alpha_p^n
    \left|\frac{E_\mathrm{c}}{E_\mathrm{g}}\right|^{|p|+2n},
\end{split}
\label{eq:Pathway}
\end{equation}
with
\begin{equation}
    \alpha_p^n=
    \binom{\frac{1}{2}(q_\mathrm{eff}+q)}{n+\frac{|p|+p}{2}}
    \binom{\frac{1}{2}(q_\mathrm{eff}-q)}{n+\frac{|p|-p}{2}}.
\end{equation}

This expression provides a photon-pathway description of the two-colour HHG process. 

Equation \ref{eq:Pathway} can be viewed as a generalized binomial expansion of the nonlinear response in the presence of a secondary field. The index $p$ labels energy exchange with the control field and determines the emitted frequency, while $n$ enumerates degenerate pathways that differ in the number of virtual photon exchanges but lead to the same final energy. This separation provides a direct link between experimentally observed spectral components and the underlying light–matter interaction pathways.

The sum over $p$ thus labels the emitted wavemixing order and therefore the energy and momentum of the emission. In the collinear geometry used here, $p$ only labels the wavemixing order. Because the generation and control wavelengths are incommensurate, the different harmonic and wavemixing orders are spectrally distinct. The sum over $n$ describes degenerate pathways that contribute to the same emitted frequency. These pathways differ by the exchange of additional pairs of control photons and therefore leave the emitted energy unchanged. Examples for the third harmonic are shown schematically in Fig. \ref{fig:pathways}e.

A central feature of the model is that consecutive degenerate pathways have opposite sign, that is, $\alpha_p^n$ and $\alpha_p^{n+1}$ contribute out of phase \cite{Vimal2023}. This can be understood physically by noting that each photon exchange corresponds to a dipole transition carrying an intrinsic phase of $-i=e^{-i\pi/2}$. Two additional control photons therefore add a phase of $-\pi$ \cite{Gerry_Knight_2004}, so neighbouring degenerate pathways interfere destructively. This directly explains why a weak control field generically suppresses the harmonic yield, while higher-order pathways can restore constructive interference and generate revivals at larger control strength.
Importantly, this phase relation is independent of microscopic material details and follows directly from the structure of dipole transitions. As a result, suppression at low control fields is a universal feature of the model, while the magnitude of the effect is governed by $q_\mathrm{eff}$.

The binomial expansion assumes only that the control field is weaker than the generation field. Implicitly, however, we also assumes that the control field does not itself modify $q_\mathrm{eff}$. This approximation breaks down at sufficiently high control strengths, where the control field can alter the underlying nonlinear response and the minimal pathway model starts to underestimate the higher-order revivals.

To calculate delay-dependent spectra, Eq. \ref{eq:Eq} was evaluated numerically in the time domain. The measured infrared spectra of the generation and control pulses were fitted with Gaussians and Fourier transformed to obtain the temporal electric fields. A quadratic spectral phase was then added to reproduce the measured pulse durations and chirp. Higher-order spectral phase terms were neglected.

To account for the absence of carrier-envelope phase stability in the experiment, the calculations were repeated for five CEP values and the resulting intensities were averaged. This averaging only affects situations in which harmonic and wavemixing signals overlap spectrally and can therefore interfere.

To include the spatial profile of the beams, the calculation was additionally repeated for different radial positions and the resulting signals were integrated. We found that this spatial averaging mainly affects cases with very low effective intensity scaling. For this reason, it was included only in the quantitative comparison shown in Fig. \ref{fig:H7_closeup}.

To incorporate emission-phase modulation, we write
\begin{equation}
\begin{split}
    \phi_e(t)=&\,\beta_\mathrm{p}\frac{|E_\mathrm{tot}(t)|^2}{\max[|E_\mathrm{c}|^2]} \\
    &+\beta_\mathrm{np}\frac{\int_{-\infty}^{t}|E_\mathrm{c}(t')|^2\,\mathrm{d}t'}{\max\left[\int_{-\infty}^{t}|E_\mathrm{c}(t')|^2\,\mathrm{d}t'\right]}.
\end{split}
\end{equation}
Here $\beta_\mathrm{p}$ and $\beta_\mathrm{np}$ denote the amplitudes of the parametric and non-parametric phase contributions, respectively. The parametric term accounts for the intrinsic intensity dependence of the emission phase \cite{Carlstrm2016,Lu2019,2025Kuzkova} and is the HHG analogue of cross-phase modulation. In the present model we approximate this contribution by a constant proportionality factor and do not explicitly retain its dependence on the relative optical phase of the two fields; for further discussion, see the Supplementary Material. The non-parametric term captures phase shifts induced by transient material excitation. In its simplest form, it is taken to scale with the accumulated control-pulse intensity, appropriate for excited-state lifetimes exceeding the pulse duration.

The non-parametric term is written as intensity dependent because multi-photon excitation is strongly wavelength dependent. For higher-order excitation pathways, a steeper scaling would be expected. We note that even when the induced material modification persists beyond the pulse overlap, spectral changes only arise while the emission phase varies during the generation pulse. Long-lived phase shifts must therefore be accessed by direct phase-sensitive measurements of the harmonics \cite{Koll2025,2019Kim,Lu2019,2025Kuzkova}.

This model is intentionally minimal. Rather than resolving all microscopic material-specific dynamics, it isolates the quantities that govern optical control in our measurements: the effective nonlinear order, which sets the pathway weights, and the emission phase, which sets their interference. This is sufficient to reproduce the main observed signatures of control, including suppression, revival, wavemixing, and delay-dependent spectral shifts.

The two phase contributions lead to distinct observable signatures in the delay-dependent spectra. 
Parametric phase modulation preserves anti-symmetric spectral features around temporal overlap, 
as it follows the instantaneous electric field. In contrast, non-parametric phase modulation, 
which depends on the accumulated excitation, breaks this anti-symmetry and leads to more symmetric 
spectral shifts. This provides a direct experimental handle to distinguish field-driven and 
material-driven contributions.

More generally, both phase terms act by modifying the relative phase of sub-cycle emission events. 
Within the pathway picture, this corresponds to a redistribution of the interference between 
degenerate pathways. In the time domain, this can be interpreted as a form of temporal phase 
matching, where the control field alters the phase relation between emission from neighbouring 
half-cycles.

\bmhead{SBE simulations}
For the simulations, we used a minimal effective two-dimensional tight-binding Hamiltonian defined on a square Bravais lattice with a lattice constant of 5.4~\AA. Each unit cell hosts two orbitals located at the same intracell position and is occupied by one electron. We build the Hamiltonian so that the minimum bandgap is in $\Gamma$ with
\begin{equation}
H_0 =
\begin{pmatrix}
\varepsilon_{11} + 2t_{11}s(\mathbf{k}) & 0 \\
0 & \varepsilon_{22} + 2t_{22}s(\mathbf{k})
\end{pmatrix},
\end{equation}
where $\varepsilon_{ii}$ are the self-energies of the orbitals, $t_{ii}$ are the first-neighbor hoppings, with $i=1,2$, and $s(\mathbf{k}) = \cos(k_x a) + \cos(k_y a)$. To get symmetric bands and the
minimum bandgap in $\Gamma$, we set a quantity $\epsilon$ such that $\varepsilon_{11} = -\varepsilon_{22} = \epsilon/2$ and $t_{11} = -t_{22} = (\epsilon - E_{bg})/8$, where $E_{bg}$ is the bandgap energy set to 2~eV. The parameter $\epsilon$ ($-\epsilon$), which we set to $\epsilon = 21.77$~eV, determines the top (bottom) of the conduction (valence) band. The optical transitions are introduced with a dipole coupling only between the two different orbitals in the same position as
$d_x = d_y = 0.1 + i0.05$ a.u..

The HHG spectra is computed using the real-time, real-space single-particle density matrix approach described in Ref. \cite{ATATA}. The electric field applied $\mathbf{E}(t)$ consists of a generation laser field with $\lambda_g=2000$~nm, intensities in the order of GW/cm$^2$ and a Gaussian envelope of 60~fs full-width at half-maximum and a control field with $\lambda_c=1400$~nm, intensities always similar or below the generation ones and a duration of 40~fs. We have used a dephasing time of $T_2=3.33$~fs, equal to half of the period of the 2000~nm laser, and a relaxation time of $T_r = 100$~fs.

\backmatter

\bmhead{Supplementary information} 
Supplementary information on the experimental setup, theory and simulations can be found at \red{Insert Link by Editor}.
%\orange{If your article has accompanying supplementary file/s please state so here. Authors reporting data from electrophoretic gels and blots should supply the full unprocessed scans for key as part of their Supplementary information. This may be requested by the editorial team/s if it is missing. Please refer to Journal-level guidance for any specific requirements.}

\bmhead{Acknowledgements}
This work has been carried out at the Advanced Research Center for Nanolithography (ARCNL), a public-private partnership of the University of Amsterdam (UvA), the Vrije Universiteit Amsterdam (VU), the Netherlands Organisation for Scientific Research (NWO), and the semiconductor equipment manufacturer ASML, and was partly financed by ‘Toeslag voor Topconsortia voor Kennis en Innovatie (TKI)’ from the Dutch Ministry of Economic Affairs and Climate Policy. This manuscript is part of a project that has received funding from the European Research Council (ERC) under the European Union’s Horizon Europe research and innovation programme (grant agreement no. 101041819, ERC Starting Grant ANACONDA) and funded P.J.v.E. and partly P.M.K. The manuscript is also part of the VIDI research programme HIMALAYA with project number VI.Vidi.223.133 financed by NWO, which partly funded P.M.K. P.M.K. acknowledges support from the Open Technology Programme (OTP) by NWO, grant no. 18703. 
A.C. and A.J.G. acknowledge support from the Talento Comunidad de Madrid Fellowship 2022-T1/IND24102, the Spanish Ministry of Science, Innovation and Universities through grant reference PID2023-146676NA-I00, and CNS2025-166331. R.E.F.S. acknowledges support from the Spanish Ministry of Science, Innovation and Universities through grants RYC2022-035373-I and CNS2024-154463.
\\
We thank Francesco Corazza for the fabrication of the silicon thin film sample. 

\bmhead{Author contributions}
P.J.v.E. and P.M.K. envisioned and conceptualized the experiments. P.J.v.E. carried out experiments, analyzed the data, and performed the analytical calculations. A.C., R.E.F.S., and A.J.G. performed the numerical simulations. P.M.K. supervised the work. P.J.v.E. and P.M.K. wrote a first version of the manuscript that was finalized by all authors.

\bmhead{Competing interests}
The authors declare no competing interests.

\bibliography{sources}

\section{Extended Data}
\begin{figure*}[p]
    \centering
    \includegraphics[width=\linewidth]{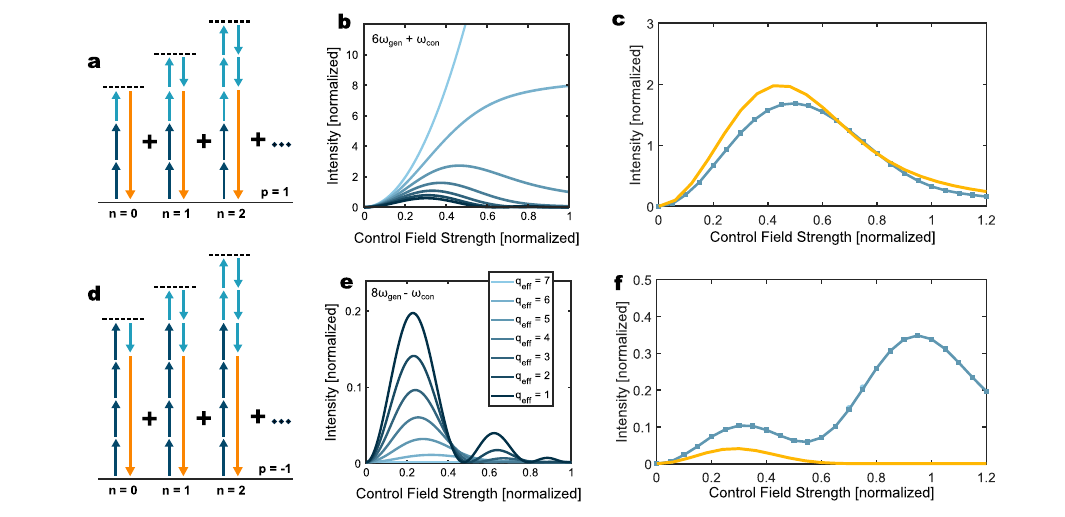}
    \caption{Schematics, calculations and simulation results for the first-order seven photon sum-frequency (a-c) and differency-frequency (d-f) wavemixing signal, corresponding to the seventh harmonic data show in Fig. \ref{fig:pathways}. In (a,d) the schematic photon pathway representation. In (b,e) the pathway calculations for various effective intensity scalings, both the graphs are normalized with respect to the seventh harmonic intensity. In (c,f) the two-band SBE simulations results for 50 GW/cm$^2$. The yellow line indicates the pathway model result for $q_\text{eff} = 4.5$. Both these graphs were normalized to the unperturbed seventh harmonic intensity. }
    \label{fig:ExtData_wavemixing}
\end{figure*}
\begin{figure*}[p]
    \centering
    \includegraphics[width=88mm]{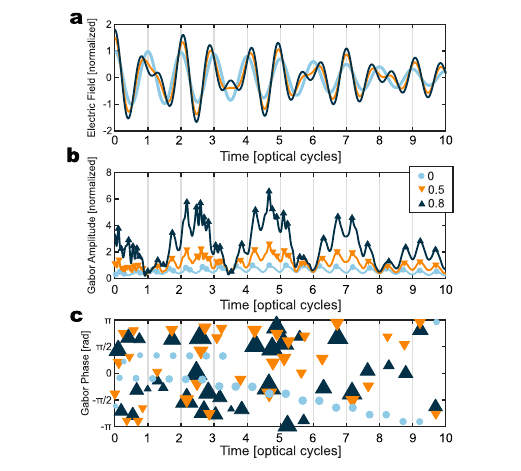}
    \caption{The Gabor transform analyses of the 50 GW/cm$^2$ SBE results for the seventh harmonic corresponding to the data shown in Fig. \ref{fig:pathways}. The legend indicates the relative control field strength. (a) shows the driving fields. In (b), the amplitude of the Gabor transform are shown where all peaks are indicated by a marker. The amplitudes are normalized to the maximum amplitude without the control field present. (c) shows the phases for each peak indicated in (b) with the marker size proportional to the related amplitude. For clarity the linear phase of the seventh harmonic has been subtracted.}
    \label{fig:ExtData_Gabor}
\end{figure*}
\begin{figure*}[p]
    \centering
    \includegraphics[width=\linewidth]{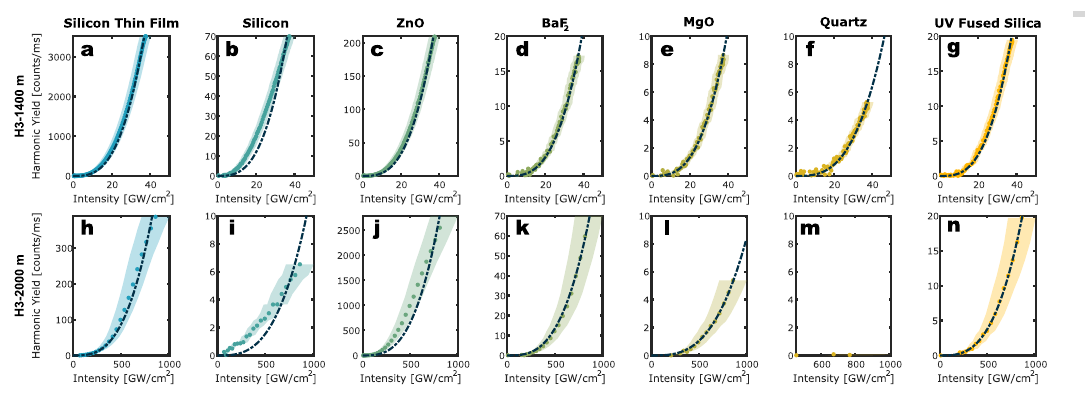}
    \caption{Experimental harmonic yield as a function of intensity for the various samples. In (a-g) for the third harmonic generated by 1400 nm and in (h-n) for the third harmonic generated by 2000 nm. The black dashed lines are exponential fits assuming perturbative intensity scaling.\\} 
    \label{fig:ExtData_IntensityScalings}
\end{figure*}
\begin{figure*}[p]
    \centering
    \includegraphics[width=88mm]{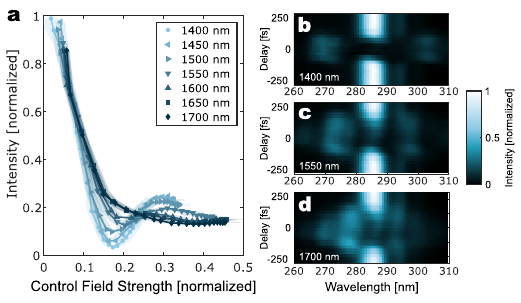}
    \caption{Experimental results of H7 generated from ZnO with a 2000 nm generation beam at 1400 GW/cm$^2$. (a)  shows the H7 harmonic intensity at the delay of 0 fs as a function of control field strength for the various control wavelengths. The shaded area indicates the uncertainty in control field strength; the uncertainty in intensity is not shown here as it is small in comparison. (b), (c) and (d)     show the delay spectra for a relative control field strength of 0.2 for various control wavelengths of 1400 nm, 1550 nm, and 1700 nm. \\
    }
    \label{fig:ExtData_WavelengthDependence}
\end{figure*}
\begin{figure*}[p]
    \centering
    \includegraphics[width=\linewidth]{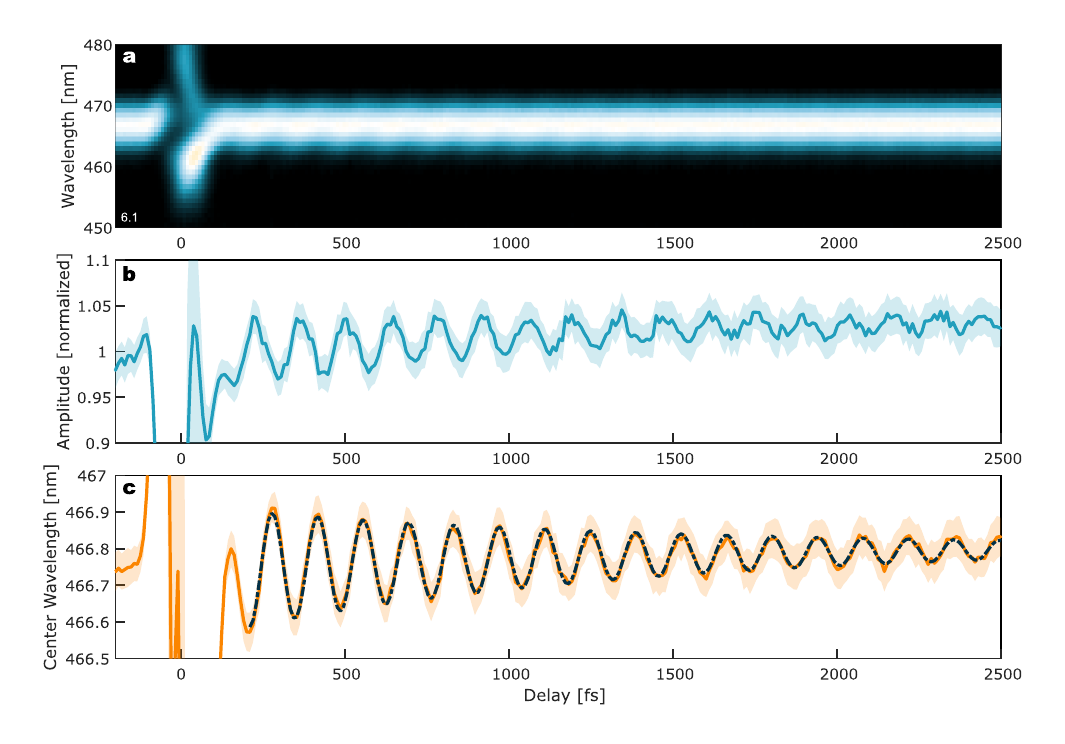}
    \caption{Observation of phonon oscillations in the third harmonic generated by 1400 nm from BaF$_2$ and modulated by 2000 nm. The 1400 nm beam has an intensity of approximately 40 GW/cm$^2$, while the 2000 nm has an intensity of about 1470  GW/cm$^2$. (a) shows the full spectra while (b) and (c) respectively show the amplitude and center wavelength resulting from a Gaussian fit. The shaded areas indicate the 95\% confidence fitting interval. The black dashed line in (c) is the fitted phonon oscillation which has an oscillation period of $138.5\pm0.3$ fs and a e$^{-1}$-life time of  $1.27\pm0.05$ ps.\\} 
    \label{fig:ExtData_Phonons}
\end{figure*}
\begin{figure*}[p]
    \centering
    \includegraphics[width=\linewidth]{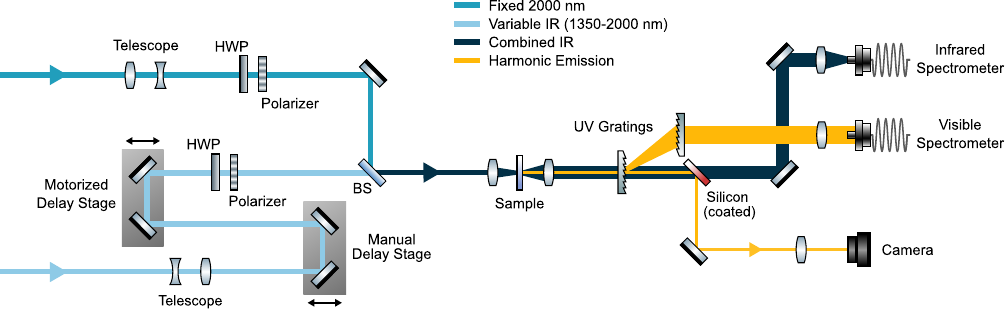}
    \caption{Schematic of the experimental setup. Some mirrors are not shown to improve readability. The half-wave plates are denoted by HWP and the beamsplitter, used here as beamcombiner, is denoted by BS.}
    \label{fig:ExtData_Schematic}
\end{figure*}
\begin{figure*}[p]
    \centering
    \includegraphics[width=\linewidth]{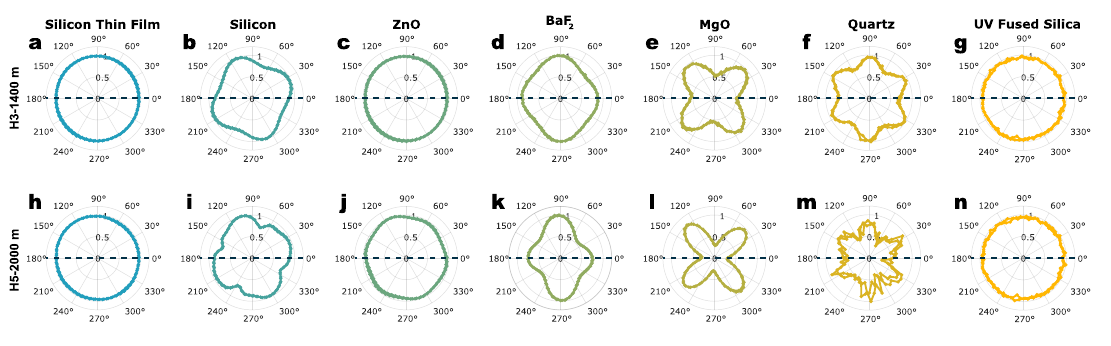}
    \caption{Experimental harmonic dependence on the laser polarization from various samples. In (a-g) for the third harmonic generated by 1400 nm and in (h-n) for the fifth harmonic generated by 2000 nm. The black dashed line indicates the polarization used during the measurements. The grating efficiency in the detection path was corrected by using the UV fused silica data.\\} 
    \label{fig:ExtData_Polarization}
\end{figure*}

%\begin{appendices}
%\section{\red{Section title of first appendix}}\label{secA1}
%\red{An appendix contains supplementary information that is not an essential part of the text itself but which may be helpful in providing a more comprehensive understanding of the research problem or it is information that is too cumbersome to be included in the body of the paper.}




%\end{appendices}

\end{document}